\documentclass[letterpaper, 10 pt, conference]{ieeeconf}  

\IEEEoverridecommandlockouts                              
\usepackage{graphics} 
\usepackage{epsfig} 
\usepackage{amsmath} 
\usepackage{amssymb}  

\usepackage{url}
\usepackage[ruled, vlined, linesnumbered]{algorithm2e}
\usepackage{verbatim} 
\usepackage{soul, color}
\usepackage{lmodern}
\usepackage{fancyhdr}
\usepackage[utf8]{inputenc}
\usepackage{fourier} 
\usepackage{array}
\usepackage{makecell}

\SetNlSty{large}{}{:}

\makeatletter

\newcommand{\Rom}[1]{\expandafter\@slowromancap\romannumeral #1@}
\makeatother

\title{\LARGE \bf
Investigations in the Displacement Current of Transverse Electromagnetic Waves and Longitudinal Interactions
}

\author{Arvin Sharma\\
New Mexico State University \\
1780 E University Ave, Las Cruces, NM 88003, United States \\
{\tt\small arvin@nmsu.edu} \\ \\
}

\begin{document}

\maketitle
\thispagestyle{plain}
\pagestyle{plain}

\begin{abstract}
The Displacement Current is a peculiar aspect of Maxwell's equation created by theoretical necessity only to be later validated through experimentation. We analyze the properties of the Displacement Current in the static condition of resulting in polarization of dielectrics and in the dynamic condition in Transverse Electromagnetic (TEM) waves. We perform this investigation to determine the role the Displacement Current has in influencing longitudinal interactions attributed to particles of light as part of the long-standing problem of physical optics known as the wave-particle duality, in addition to exploration of the theories of faster than light propagation in theoretical longitudinal modes of electric transmission attributed to the Displacement Current.
\end{abstract}


\section{INTRODUCTION}

James Clerk Maxwell's investigations in electrodynamic theory led him to formulate Ampère's Circuital Law which related the circulation of magnetism to the conduction current. This creates an inconsistency in which a break in the circuit, or two capacitor plates separated by a dielectric, exists. In order to resolve this ambiguity, since direct current flows until the capacitor is charged, the displacement current, $\frac{\partial{D}}{\partial{t}}$, as a secondary component flowing between the plates of the capacitor, has been added:\par
\[\oint{B\cdot dl} = \int{\mu_0(J + \frac{\partial{D}}{\partial{t}} )\cdot dS}\]
By applying Stoke's theorem to the above formulation of Ampère's Circuital Law, we arrive at one of the four Maxwell equations:
\[\nabla \times B = \mu_0(J + \frac{\partial{D}}{\partial{t}})
\]

The displacement current is an inherently wireless phenomenon, due to the lack of a conductor as in the case of the conduction current. However, it is not immaterial because it requires a dielectric, be it a physical material or the vacuum, as the medium of transmission. In the case of the vacuum, historical theories relied upon a mechanical ether in order to explain transverse electromagnetic wave propagation \cite{c1}. While mechanical theories of the ether have been set aside in favor of the wave-particle duality which explains certain quantum phenomena, its usefulness in providing a clear mechanical analog for electromagnetic phenomena has led to its persistence, in the form of a mathematical field, in applied science and engineering \cite{c2}.\par

Nikola Tesla in patent No. 787,412 describes a communication system with propagates at speeds he claims to be faster than light \cite{c3}. In traditional electromagnetic theory, this is not possible due to the velocity of light being defined as constants for the vacuum \cite{c4}, found in the wave equation which describes the propagation of TEM waves where $u=\frac{1}{\sqrt{\mu\epsilon}}$ \cite{c7}:\par
\[\nabla^2{E} - \frac{1}{u^2}\frac{\partial^2{E}}{\partial{t}^2} = 0\]
\[\nabla^2{H} - \frac{1}{u^2}\frac{\partial^2{H}}{\partial{t}^2} = 0\]
Should Nikola Tesla's system operate outside of traditional electromagnetics, and we allow flexibility in the definitions for the limits of reality, we may speculate that in an alternate mode of electromagnetic wave propagation, Tesla's measurements are possible. In this case, we will investigate the possibility for Longitudinal Electromagnetic (LEM) wave propagation as initiated by Maxwell's displacement current, as offering one such explanation, and extend this into an investigation for the possibility of an explanation of phenomena currently attributed to the particulate nature of light.

\section{WAVE PROPAGATION IN MEDIA}

A wave is defined as a disturbance exhibiting periodicity. We concern ourselves with transverse waves, that is waves which exhibit the property of motion transverse to the direction of propagation. Longitudinal waves oscillate along the direction or axis of propagation. For simplicity we are concerned with the class of waves exhibiting simple harmonic motion (SHM):\par
\[y = Asin(\omega t)\]
For ideal transverse waves, each point in the oscillation moves with simple harmonic motion along the axis perpendicular to the direction of propagation. In classical electromagnetics, this is the electric and magnetic field at right angles oscillating with SHM along the propagation path. As we shall see with transverse surface waves in media, there exists longitudinal components due to broad side pressures from adjacent particles causing the particle to oscillate elliptically.\par 
Transverse waves in fluids such as water exhibit longitudinal components as a result of the pressure gradient formed within the medium and the force of gravity. When modeling ocean waves, Trochoidal waves are used as opposed to Sinosoidal \cite{c5}. In this formulation, each particle has velocity components in both the transverse and longitudinal directions. In which $(x_0,y_0)$ is the location of the particle at rest and $k$ the wave number:\par
\[x = x_0 + Ae^{ky_0}sin(\omega t-kx_0)\]
\[y = y_0 + Ae^{ky_0}cos(\omega t-kx_0)\]
For solid medium, transverse surface waves propagate as Raleigh waves in which each point on the object exhibit elliptical oscillations \cite{c6}.\par

In summary, we see a longitudinal component exists within liquids and solids when a transverse surface wave is propagated through the media. This complicates the mechanical notion of the ether for in traditional electromagnetics no longitudinal component is present in the wave equation for propagation. Knowing the transverse nature of electromagnetic waves, and the presence of longitudinal waves in transverse oscillations within media, we shall now investigate the Displacement Current, Nikola Tesla, and the possibility of LEM waves through dielectric media.\par

\section{DISPLACEMENT CURRENT}
Displacement current is defined to be the time variation of the Electric Flux Density $D$, giving rise to the curl of the Magnetic Field Intensity $H$, where in free space:
\[\nabla \times H = \frac{\partial{D}}{\partial{t}}
\]\cite{c7}
Which has the physical meaning of the time variation of electric flux over a particular area of space, that is, the time variation of electric flux density. Electric flux Density, $D$ is defined from Gauss' Law $\int{D}\cdot dS = \Psi$, where $D = \frac{\Psi}{S}$ and $\Psi = Q$ in Coulombs, as the quantity of electric flux through an area in space, and is hypothesized by Maxwell to be a physical entity in the all-encompassing medium \cite{c8}.\par
The character of the propagation of power through the Displacement Current is distinct from Conduction Current, which is defined as the movement of electrons through a region of space. In the case of the Displacement Current, it is the change in polarization of the dielectric medium transmitting the power. The reason LEM waves are possible with Displacement Current is clear with a mechanical analogy. If we have a rope in which a pull in one end is propagated to the end of the other side, the force must have propagated longitudinally, since no transverse motion is found across the rope. In a similar manner, an increase in the electric field increases the polarization of the medium, which for media that is linear and isotropic with electric susceptibility $\chi_e$ we find the following relation \cite{c7}:
\[P = \epsilon_0\chi_e E\]
Such that the total polarization is directly proportional to the Electric field intensity. Note that polarization is defined as the sum of induced dipoles, that is:
\[P = \lim_{\Delta{v} \to 0}\frac{\sum_{k=1}^{n\Delta{v}} p_k}{\Delta{v}}\]
Where $p_k$ is an individual dipole moment in the material. This means an increase in P is an increase in the dipole moment, where the molecule is even more polarized. Since the displacement current originates in one side of the capacitor plate, on the sending end, this creates a chain reaction through the medium as each dipole reorients and polarizes to the new field, which by necessity is longitudinal as each dipole is oriented in the direction of the field. This clear analog has been demonstrated, albeit outside of the traditional academic institutions, by Eric P. Dollard, electrical engineer and experimentalist, in the 1980s, who to the authors knowledge is credited with discovering the possible longitudinal modes for electromagnetic waves. \cite{c9}
\subsection{Applications in Electromagnetic Theory}
Using classical electromagnetic theory, in which a plane wave of amplitude $E_0$ propagates in the $+z$ direction with the relation:
\[E(z) = E_0e^{-jkz}\]
One find the theory completely impenetrable for a longitudinal mode along the path of propagation. Therefore, we must turn to the transmission apparatus itself before the TEM wave is initiated to find the possibility for a longitudinal mode to exist. In particular, the elevation capacity or antenna, such as the one incorporated in Nikola Tesla's Wardenclyff experiment. \cite{c10}.\par
In this apparatus, a large elevative capacity, dome shaped is joined with a deep underground network of metallic structures. The theory of communication involves electric fields linking the elevative and underground structures together. In this configuration, each antenna acts as one side of the capacitor, with one half of each side underground. Should the ground be conductive, which has been demonstrated with single wire transmission systems using the ground as the return \cite{c11}, then it is between the elevative capacities and the conductive earth in which power transmission takes place. It is hypothesized in this configuration, for Tesla's claim of faster than light transmission to take place, this must involve something different from TEM, such as the LEM mode. If the ground network interacts with the geology of the ground to form consecutive pairs of capacitances depending on the mineral composition, a longitudinal train of electric polarization has the capability of taking place in the dielectric materials of the earth, propagating in all directions to the destination, depicting by Tesla's own diagram\cite{c12}:\par

\includegraphics[scale=0.15]{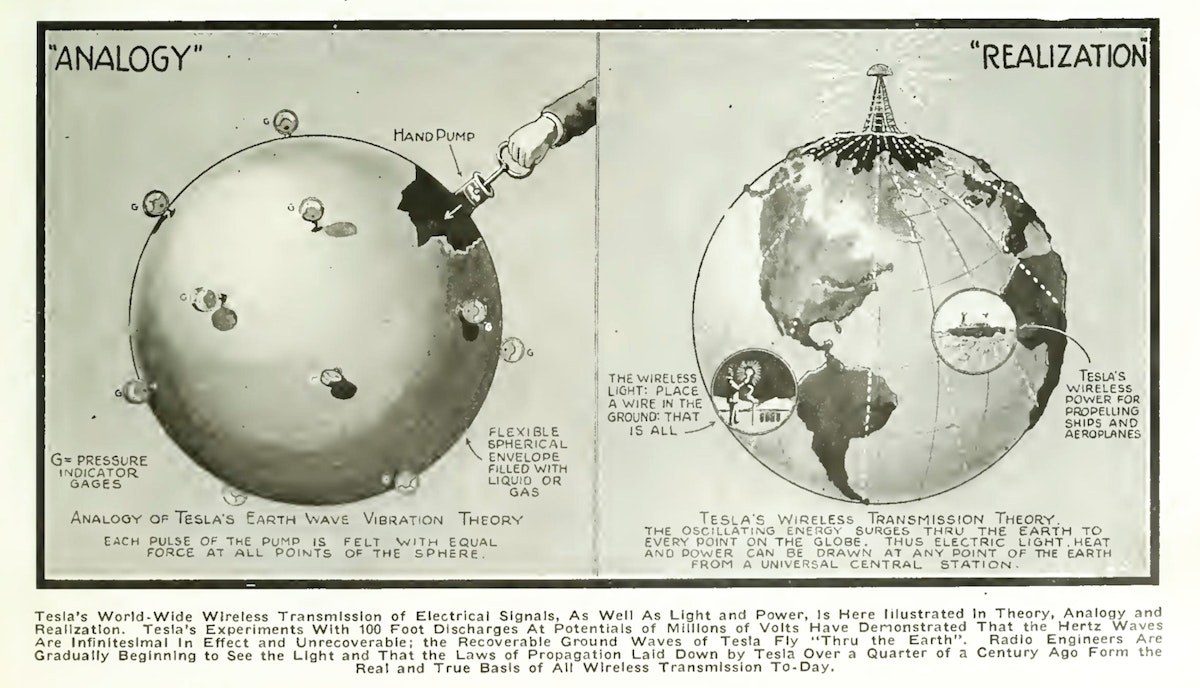}\\
Where the transmitter acts as a pump sending longitudinal waves tellurically to each receiver. To the author's knowledge the speed of this propagation which is purely electric is only determined thus far by Nikola Tesla in his patent \cite{c3} and is an open space for experimenters and researchers to investigate.
\section{CONCLUSION}

While it is not likely a longitudinal mode exists in TEM waves where the Electric field is transverse to the direction of propagation, there exists the possibility for other modes of electric power transmission utilizing the Displacement Current and dielectric polarization as the primary transmission medium, agnostic of the magnetic field. In addition, there exists the area of exploration in the interaction of TEM waves with matter, in which at a very local level to a material particle polarization with the nearby field has reason for taking place. The interaction of the transverse polarization of materials from the transverse Electric field opens the space to investigate fundamental optics interactions such as Huygens' Principal and the effects of Diffraction from the standpoint of dipole interactions from polarized media, coloring a perspective in wave optics harmonious with the classical formulation of electromagnetics.

\addtolength{\textheight}{-12cm}   



\end{document}